\documentstyle[preprint,aps,prl]{revtex}
\begin{document}
\draft
\preprint{accepted for publication in Phys.~Rev.~Lett.}
\title{
Microscopic Theory of Quantum-Transport Phenomena in Mesoscopic Systems: \\
A Monte Carlo Approach
}
\author{Fausto Rossi}
\address{
Istituto Nazionale per la Fisica della Materia (INFM) and 
Dipartimento di Fisica, Universit\`a di Modena, I-41100 Modena, Italy 
}
\author{Aldo Di Carlo and Paolo Lugli} 
\address{
INFM and Dipartimento di Ingegneria Elettronica, 
Universit\`a di Roma ``Tor Vergata'',
I-00133 Roma, Italy
}
\date{\today}
\maketitle
\begin{abstract}

A theoretical investigation of quantum-transport phenomena in mesoscopic 
systems is presented. In particular, a generalization to ``open systems'' 
of the well-known semiconductor Bloch equations is proposed. 
The presence of spatial 
boundary conditions manifest itself through 
self-energy corrections and additional source terms in the kinetic 
equations, whose form is suitable for a solution via 
a generalized Monte Carlo simulation.
The proposed approach is applied to the study of quantum-transport 
phenomena in double-barrier structures as well as in superlattices, 
showing a strong interplay between phase coherence and relaxation.

\end{abstract}
\narrowtext

The Monte Carlo (MC) method, which has been applied for more than
25 years for calculation of semiclassical
charge transport in semiconductors,
is the most powerful numerical tool for 
microelectronics device simulation \cite{MC}. 
However, present-day
technology pushes device dimensions toward limits where the
traditional semiclassical transport theory can no longer be applied,
and a more rigorous quantum transport theory is 
required \cite{QT}. 
To this end, various quantum-kinetic formulations of 
charge transport have been proposed, based on Green's function \cite{GF} 
or Wigner-function \cite{WF} approaches.
While such quantum-mechanical formalisms allow for a rigorous 
treatment of phase coherence, they typically describe energy-relaxation and
dephasing processes via purely phenomenological models.
A full quantum-mechanical 
simulation scheme for the analysis of transient-transport phenomena in the 
presence of carrier-phonon interaction has also been  proposed \cite{QMC}. 
However, due to the huge amount of computation
required, its applicability is still limited to 
short time-scales and extremely simplified situations.
As a result,
despite many efforts and despite the unquestionable intellectual 
progress associated with the study of these quantum-kinetic formulations, 
their application to realistic devices in the presence of a 
strong scattering dynamics is still an open problem.
Recent results by Datta, Lake, and co-workers seem to be rather promising
\cite{Datta}.
However, their steady-state Green's function formulation cannot be applied 
to the analysis of time-dependent non equilibrium phenomena, 
which play a crucial role in modern optoelectronic devices.

In this letter we propose a generalized MC approach for the analysis 
of hot-carrier transport and relaxation phenomena 
in quantum devices.
The method is based on a MC
solution of the set of kinetic equations governing the time evolution of 
the single-particle density matrix. 
Our approach can be regarded as an extension to open systems
of the generalized MC method recently proposed for the analysis of the 
coupled coherent and incoherent carrier dynamics in photoexcited 
semiconductors \cite{GMC}.
Compared to more academic quantum-kinetic approaches 
\cite{GF,WF,QMC} ---whose 
application is often limited to highly simplified physical models and 
conditions---, 
the proposed simulation scheme allows to 
maintain all the well known advantages of the 
MC method in describing a large variety of scattering 
mechanisms on a microscopic level \cite{MC}.

In order to properly describe carrier-transport phenomena in mesoscopic 
structures, an electron-phonon system can be considered, whose 
Hamiltonian can be schematically written as 
$
{\bf H} = {\bf H}_\circ + {\bf H}'
$.
Here, the single-particle term ${\bf H}_\circ$ includes the  phonon
and free-carrier   Hamiltonians as well as the potential profile 
(including possible external  fields), while the 
many-body contribution ${\bf H}'$ accounts for all possible interaction 
mechanisms, e.g. carrier-carrier and carrier-phonon coupling.
By denoting with 
$
\phi_\alpha({\bf r}) = \langle {\bf r} \vert \alpha \rangle
$
the wavefunctions of the single-particle states $\alpha$ \cite{note1}
and  with $\epsilon_\alpha$  the corresponding 
energies,  the 
 equation of motion for the single-particle 
density matrix $\rho$ \cite{note2} in this $\alpha$-representation 
can be schematically written as:
\begin{equation}\label{SBE-total}
\frac{d}{dt} \rho_{\alpha\beta} = 
\frac{d}{dt} \rho_{\alpha\beta}\Bigl|_{{\bf H}_\circ} + 
\frac{d}{dt} \rho_{\alpha\beta}\Bigl|_{{\bf H}'}\ .
\end{equation}
The time evolution induced by the single-particle Hamiltonian ${\bf H}_\circ$
can be evaluated exactly. 
In contrast, the contribution due to the many-body Hamiltonian 
${\bf H}'$ involves phonon-assisted as well as higher-order density-matrices; 
thus in order to ``close'' our set of 
equations approximations are needed.
In particular, as described in \cite{Kuhn97}, the ``mean-field'' 
approximation together with the Markov 
limit leads to a closed set of equations still local in time.
Within such approximation scheme, the equations of motion in 
(\ref{SBE-total}) can be written as
\begin{equation}\label{SBE}
\frac{d}{dt} \rho_{\alpha\beta} = 
\sum_{\alpha'\beta'} L_{\alpha\beta,\alpha'\beta'} \rho_{\alpha'\beta'}
\end{equation}
with 
\begin{equation}\label{L}
L_{\alpha\beta,\alpha'\beta'} =
{1 \over i\hbar} \left(\epsilon_\alpha-\epsilon_\beta\right) 
\delta_{\alpha\beta,\alpha'\beta'} + \Gamma_{\alpha\beta,\alpha'\beta'}\ .
\end{equation}
Here, the two terms correspond   to the separation in (\ref{SBE-total}).
The explicit form of the scattering tensor $\Gamma$ involves the 
microscopic in- and out-scattering rates for the various interaction 
mechanisms \cite{note3}.
The above kinetic equations can be regarded as a multiband generalization 
of the well known 
semiconductor Bloch equations (SBE) \cite{SBE}.

The analysis presented so far is typical of a so-called ``closed'' system, 
i.e. a physical system defined over the whole coordinate space. 
However, this is not the case of interest for the study of 
quantum-transport phenomena in mesoscopic devices, where the 
properties of the carrier subsystem are strongly influenced by the spatial 
boundaries with the external environment. 
This requires a  real-space description, which can be 
obtained in terms of  the phase-space formulation of quantum mechanics 
originally proposed by Wigner \cite{Wigner} and generalized to solids in the 
pioneering paper by Buot \cite{Buot}.
In our case, this corresponds to introduce the following unitary 
transformation $u$ connecting our $\alpha\beta$ representation to the 
desired phase-space ${\bf r,k}$:
\begin{equation}\label{u}
u_{\alpha\beta}({\bf r,k}) = (2\pi)^{-{3\over 2}} \int d{\bf r}' 
\phi^{ }_\alpha\left({\bf r}+{1\over 2}{\bf r}'\right) 
e^{-i{\bf k \cdot r}'}
\phi^*_\beta\left({\bf r}-{1\over 2}{\bf r}'\right)\ .
\end{equation}
By applying the above Weyl transform to the single-particle density matrix 
$\rho$, we obtain the so-called Wigner function \cite{WF}:
\begin{equation}\label{WF}
f^W({\bf r,k}) = \sum_{\alpha\beta} 
\rho_{\alpha\beta} u_{\alpha\beta}({\bf r,k})\ .
\end{equation}
For a closed system, the Wigner function $f^W$ is defined for 
any value of the real-space coordinate ${\bf r}$ and its time evolution is 
fully determined by its initial condition.
In contrast, for an open system $f^W$ is defined only within a 
given region $\Omega$ of interest and its time evolution is determined by 
the initial condition  inside such region as well as by its 
values $f^W_b$
on the boundary ${\bf r}_b$ of the domain at any time 
$t' > t_\circ$. 
More specifically, by applying the Green's function theory to the equation 
of motion for $f^W$ ---which is obtained by applying to Eq.~(\ref{SBE}) the
Weyl-Wigner transform (\ref{WF})---
we get:
\begin{eqnarray}\label{open1}
f^W({\bf r,k};t)
&=& \int_\Omega d{\bf r}' \int d{\bf k}' 
G({\bf r,k}; {\bf r',k'};t-t_\circ) f^W({\bf r',k'};t_\circ) \nonumber \\
& & +
\int d{\bf r}_b \int d{\bf k}' \int_{t_\circ}^t dt'
G({\bf r,k}; {\bf r}_b,{\bf k}'; t-t')
f^W_b({\bf r}_b,{\bf k}',t') v({\bf k}')\ ,
\end{eqnarray}
where
\begin{equation}\label{EO}
G({\bf r,k}; {\bf r',k'}; \tau) = \sum_{\alpha\beta,\alpha'\beta'} 
u^{ }_{\alpha\beta}({\bf r,k}) 
\left[e^{L\tau}\right]_{\alpha\beta,\alpha'\beta'}
u^*_{\alpha'\beta'}({\bf r',k'})
\end{equation}
is the evolution operator, while $v({\bf k})$ is the 
component of the carrier group velocity normal to the boundary surface.
We clearly see that the value of $f^W$ is obtained from the 
propagation of the initial condition $f^W(t_\circ)$ inside the domain $\Omega$ 
plus the propagation of the boundary values $f^W_b$ from the points of the 
surface at any time $t'$ to the point ${\bf r,k}$ of interest.

Given the above Wigner formulation for open systems, we now 
introduce a corresponding density-matrix description via the 
following ``inverse'' Weyl-Wigner transform \cite{note4}:
\begin{equation}\label{inv}
\overline{\rho}_{\alpha\beta} = \int_\Omega d{\bf r} \int 
d{\bf k} 
u^*_{\alpha\beta}({\bf r,k}) f^W({\bf r,k})\ .
\end{equation}
By applying the above transformation to Eq.~(\ref{open1}) and then performing 
its time derivative, we finally obtain:
\begin{equation}\label{SBE-open}
\frac{d}{dt} \overline{\rho}_{\alpha\beta} = 
\sum_{\alpha'\beta'} 
\overline{L}_{\alpha\beta,\alpha'\beta'} \overline{\rho}_{\alpha'\beta'} 
+ \overline{S}_{\alpha\beta}\ ,
\end{equation}
where 
$\overline{L} = U^{ } L U^{-1}$
is the Liouville tensor (\ref{L}) 
``dressed'' by the transformation 
\begin{equation}\label{U}
U_{\alpha\beta,\alpha'\beta'} = \int_\Omega d{\bf r} \int d{\bf k} 
u^*_{\alpha\beta}({\bf r,k})
u^{ }_{\alpha'\beta'}({\bf r,k})\ ,
\end{equation}
while 
\begin{equation}\label{SBE-source}
\overline{S}_{\alpha\beta} = \sum_{\alpha'\beta'} U_{\alpha\beta,\alpha'\beta'}
\int d{\bf r}_b \int d{\bf k} 
u^*_{\alpha'\beta'}({\bf r}_b,{\bf k})
v({\bf k})
f^W_b({\bf r}_b,{\bf k}) 
\end{equation}
is a source term induced by our spatial boundary conditions.

Equation (\ref{SBE-open}) is the desired generalization to the case of
open systems of the standard SBE in Eq.~(\ref{SBE}). 
In addition to the source term in Eq.~(\ref{SBE-source}), 
the presence of spatial boundary conditions induces modifications 
on the Liouville operator $L$ of the system via the transformation $U$ in 
Eq.~(\ref{U}).  

The generalized SBE (\ref{SBE-open}) can 
be still regarded as the result of a coherent single-particle dynamics
plus incoherent many-body contributions [see Eq.~(\ref{SBE-total})]. 
Therefore, they can be solved by 
means of the same MC simulation scheme described in \onlinecite{GMC}.
The method is based on a time-step separation between coherent and 
incoherent dynamics: 
The former accounts in a rigorous way for all quantum 
phenomena induced by the potential profile of the device as well as for 
the proper boundary conditions. 
The latter, described within the basis 
given by the eigenstates $\alpha$ of the potential profile \cite{note5}, 
accounts for all the relevant scattering mechanisms by means of a 
conventional ``ensemble'' MC simulation \cite{GMC}. 

In order to illustrate the power and flexibility of the proposed theoretical 
approach, we have simulated quantum-transport phenomena in rather different 
physical systems, namely 
double-barrier structures and superlattices.
Since we are interested in low temperature and low carrier density conditions,
only optical-phonon scattering has been considered.
We have first carried out  the simulation of an electron
wavepacket entering the double-barrier structure 
\cite{note6} of a GaAs/AlGaAs resonant tunneling diode (RTD). 
Figure \ref{fig1} shows the time evolution of the wavepacket in the absence
of scattering as a function of position (a) 
and energy (b). 
It is easy to recognize the well-established resonance scenario typical of 
any purely coherent dynamics:
as the wavepacket enters our RTD 
structure, a part of it is 
transmitted and a part is reflected [see Fig.~\ref{fig1}(a)].
Since in this simulation scattering is not included, the wavepacket central
energy is conserved, i.e. no energy relaxation occurs [see Fig.~\ref{fig1}(b)].
On the contrary, in the presence of incoherent scattering processes 
the resonance dynamics of Fig.~\ref{fig1}(a) is strongly modified by the 
scattering itself, as shown in Fig.~\ref{fig2}(a).
In particular, the presence of phase-breaking scattering processes is 
found to reduce both the interference peaks and the transmitted wavepacket. 
This is confirmed by the corresponding energy distribution in 
Fig.~\ref{fig2}(b), where we clearly recognize the granular nature of the 
dissipation process through the formation of optical-phonon replica. 
This is the fingerprint of
any full microscopic treatment of energy relaxation, thus confirming the 
microscopic nature of our quantum-mechanical simulation, in contrast with
all previous phenomenological approaches.

As a second testbed for our generalized MC approach, we have also simulated 
electrically injected Bloch oscillations in superlattices (SLs). 
The system under investigation consists of a biased GaAs/AlGaAs SL 
\cite{note7} surrounded by two semi-infinite GaAs regions. 
In our simulated experiment an electron wavepacket is injected from the left 
contact (GaAs region) 
into the SL region [see Fig.~\ref{fig3}]. 
Figures \ref{fig3} and \ref{fig4} illustrate the time evolution 
of the wavepacket with and without 
scattering, respectively.
When the wavepacket reaches the SL structure most of it gets reflected 
backwards, while some portion of it tunnels into the Wannier-Stark ladder 
associated with the SL and starts to oscillate at a frequency of about 
$3.5$\,THz. 
Every time the packet reaches the boundary
of the oscillation region a part of it gets transmitted via Zener tunneling. 
We should notice, however,
that such Zener processes do not destroy the Bloch-oscillation dynamics, 
but simply reduce the 
charge density within the SL region. 
Indeed, in the scattering-free case [see Fig.~\ref{fig3}] the 
Bloch oscillations are found to persist on a picosecond time-scale.
In contrast, once scattering mechanisms are considered 
[see Fig.~\ref{fig4}], the phonon-induced dephasing drastically 
reduces their lifetime. 
As for the case of the RTD, the carrier dynamics is the result 
of a strong interplay between phase coherence and relaxation.

In conclusion, we have proposed a generalization to open systems of the 
well-known SBE.
This approach allows for a proper description of the strong 
coupling between coherent and 
incoherent dynamics. 
Indeed, our simulated experiments clearly show the
failure of any purely coherent or incoherent 
approach in describing typical quantum-transport phenomena in semiconductor 
nanostructures.

Contrary to 
all previous quantum-transport investigations, 
the proposed theoretical scheme allows to 
fully recover ---and benefit from--- the unquestionable advantages of the 
semiclassical Monte Carlo simulation, thus opening the way to the 
theoretical modelling of realistic quantum devices.
\medskip\par\noindent
We are grateful to Carlo Jacoboni, Tilmann Kuhn, Roger Lake, and 
Massimo Fischetti 
for stimulating and fruitful discussions. 
This work was supported in part by the EC Commission through the TMR Network 
``ULTRAFAST''. 
\begin{figure}
\caption{\label{fig1} 
Carrier density at different times as a function of position (a) 
and energy (b)  
corresponding to an electron wavepacket injected into a RTD structure
in the absence of scattering processes
(the two barriers are schematically depicted as black vertical lines). 
}
\end{figure}
\begin{figure}
\caption{\label{fig2} 
Same as in Fig.~\protect\ref{fig1}\protect, but in the presence of scattering 
processes (see text).
}
\end{figure}
\begin{figure}
\caption{\label{fig3} 
Contour plot of the charge density corresponding to a wavepacket 
electrically injected into a finite SL region
(marked with vertical white lines)
in the absence of scattering processes.
}
\end{figure}
\begin{figure}
\caption{\label{fig4} 
Same as in Fig.~\protect\ref{fig3}\protect, but in the presence of scattering 
processes (see text).
}
\end{figure}

\begin{references}
\bibitem{note2}
This is defined as the average value of two creation and destruction 
operators, i.e. 
$\rho_{\alpha\beta} = \langle a^\dagger_\beta a^{ }_\alpha \rangle$.
Its diagonal elements correspond to the 
usual distribution functions 
of the semiclassical Boltzmann theory \protect\cite{MC}\protect
while the off-diagonal terms ($\alpha\ne\beta$) describe 
the degree of quantum-mechanical phase coherence 
between states 
$\alpha$ and $\beta$.
\bibitem{note3}
The resulting scattering model is a generalization of the Boltzmann 
collision term \protect\cite{MC}\protect. 
Its in- minus out-scattering structure allows for a 
proper description of low-dephasing regimes typical of quasielastic processes
\protect\cite{GMC,Kuhn97}\protect.
\bibitem{note4}
Here, the word ``inverse'' refers to the fact that it allows to go back 
to the original density-matrix picture. This is clearly not the inverse of the 
transformation $u$ in Eq.~(\protect\ref{u}\protect): 
the effective density matrix 
$\overline{\rho}$ describes the region $\Omega$ of interest only, and is 
conceptually different from the original density matrix $\rho$.
\bibitem{note6}
The barrier height is $0.5$\,eV while the barrier width 
and separation are, respectively, $20$ and 
$60$\,\AA.
\bibitem{note7}
The SL structure is made by $21$ periods of 
a $8$\,\AA\ AlGaAs barrier and $57$\,\AA\ GaAs well unit cell 
with a barrier height of $0.4$\,eV.
\end{references}
\end{document}